\documentclass[11pt,debug, twoside]{article}
\usepackage{asp2014}

\aspSuppressVolSlug
\resetcounters

\begin{document}

\title{Best licensing practices}

\author{Y.~G.~Grange$^1$, T.~J\"{u}rges$^1$, J.~Schnabel$^2$, N.~P.~F.~Lorente$^3$, and M.~F\"{u}{\ss}ling$^4$}
\affil{$^1$ASTRON, the Netherlands Institute for Radio Astronomy, Oude Hoogeveensedijk 4, 7991 PD Dwingeloo, The Netherlands; \email{grange@astron.nl}, \email{jurges@astron.nl}}
\affil{$^2$Friedrich-Alexander-Universit\"{a}t Erlangen-N\"{u}rnberg, Erlangen Centre for Astroparticle Physics, Erwin-Rommel-Str. 1, 91058 Erlangen, Germany}
\affil{$^3$Australian Astronomical Observatory, 105 Delhi Road, North Ryde, NSW 2113, Australia}
\affil{$^4$DESY, D-15738 Zeuthen, Germany}

\paperauthor{Y.~G.~Grange}{grange@astron.nl}{0000-0001-5125-9539}{ASTRON}{}{Dwingeloo}{}{7991PD}{the Netherlands}
\paperauthor{T. J\"urges}{jurges@astron.nl}{0000-0002-3993-7576}{ASTRON}{}{Dwingeloo}{}{7991PD}{The Netherlands}
\paperauthor{J.~Schnabel}{jutta.schnabel@fau.de}{0000-0003-1233-7738}{Friedrich-Alexander-Universit\"{a}t}{Centre for Astroparticle Physics}{Erlangen}{}{91058}{Germany}
\paperauthor{M.~F\"{u}{\ss}ling}{matthias.fuessling@cta-observatory.org}{}{DESY}{}{Zeuthen}{}{15738}{Germany}
\paperauthor{Nuria~P.~F.~Lorente}{Nuria.Lorente@mq.edu.au}{0000-0003-0450-4807}{Australian Astronomical Observatory}{}{North Ryde}{NSW}{2113}{Australia}



  
\begin{abstract}


The principle that research output should be open has, in recent years, been increasingly applied to data and software. Licensing is a key aspect to openness. Navigating the landscape of open source licenses can lead to complex discussions.

During ADASS XXIX in 2019 it became clear that several groups worldwide are working on formalising the licensing of software and other digital assets. In this article, we summarise a discussion we had at ADASS~XXX on the application of licenses to astronomical scientific software, and summarise the questionnaire we distributed in preparation. We conclude that this topic is considered relevant and interesting by many members of our community, and that it should be pursued further.
\end{abstract}

\section{Introduction}
The application of the FAIR standards, as defined in \citep{2016_Wilkinson_FAIR}, to software makes the application of a license aiming to ensure reusability a core necessity. 

A \textit{license} defines the terms and conditions under which source code can be reused. If no license is applied, the reuse of source code is generally strongly limited due to copyright protection. Applying a license, and more specifically an open source license\footnote{In this paper, we use the term ``Open Source'' and ``Open Source license'' as they are defined at  \url{https://opensource.org/docs/osd}}, facilitates making the source code reusable.

There are roughly three types of open source license:
\begin{description}
\item[Copyleft]   (e.g. GPL) -- Allows reuse of code as long as any derived product is released under similar conditions. This extends to source code linking to a compiled library licensed under GPL. 
\item[Lesser copyleft] (e.g. LGPL) -- Applies to the code of the library itself, but not to source code linked to it.
\item[permissive] (e.g. BSD, Apache 2.0) -- Allows for reuse without many limitations. Adapted code can be used for any purposed and released under any license.
\end{description}

ASTRON's Open Source Committee presented their path towards an Open Source Policy during ADASS XXIX, 2019 \citep{2019arXiv191100534G}. In 2020, the \href{http://www.escape2020.eu}{ESCAPE project} began discussion about issues concerning software and data licensing. The apparently broad interest in this topic led to our proposal for a BOF session at ADASS~XXX. Here, we summarise the responses to a questionnaire that we distributed among the participants prior to the BoF, and give a recapitulation of the discussion held during the BoF session. 

\section{Questionnaire on use of licenses}
Participants to the conference were asked to fill a questionnaire before or during the session. In total, 51 people responded. The questionnaire consisted of three multiple-choice questions, which we discuss in the following sections. The other two questions were used to provide input on the discussion, as summarised in Sec. \ref{sec:disc}, and to point out interesting resources for more information.  We have added some of them to the extra material mentioned in Sec. \ref{sec:forward}. 

\subsection{What is the default license of your institute/collaboration w.r.t. to software?}

\articlefiguretwo{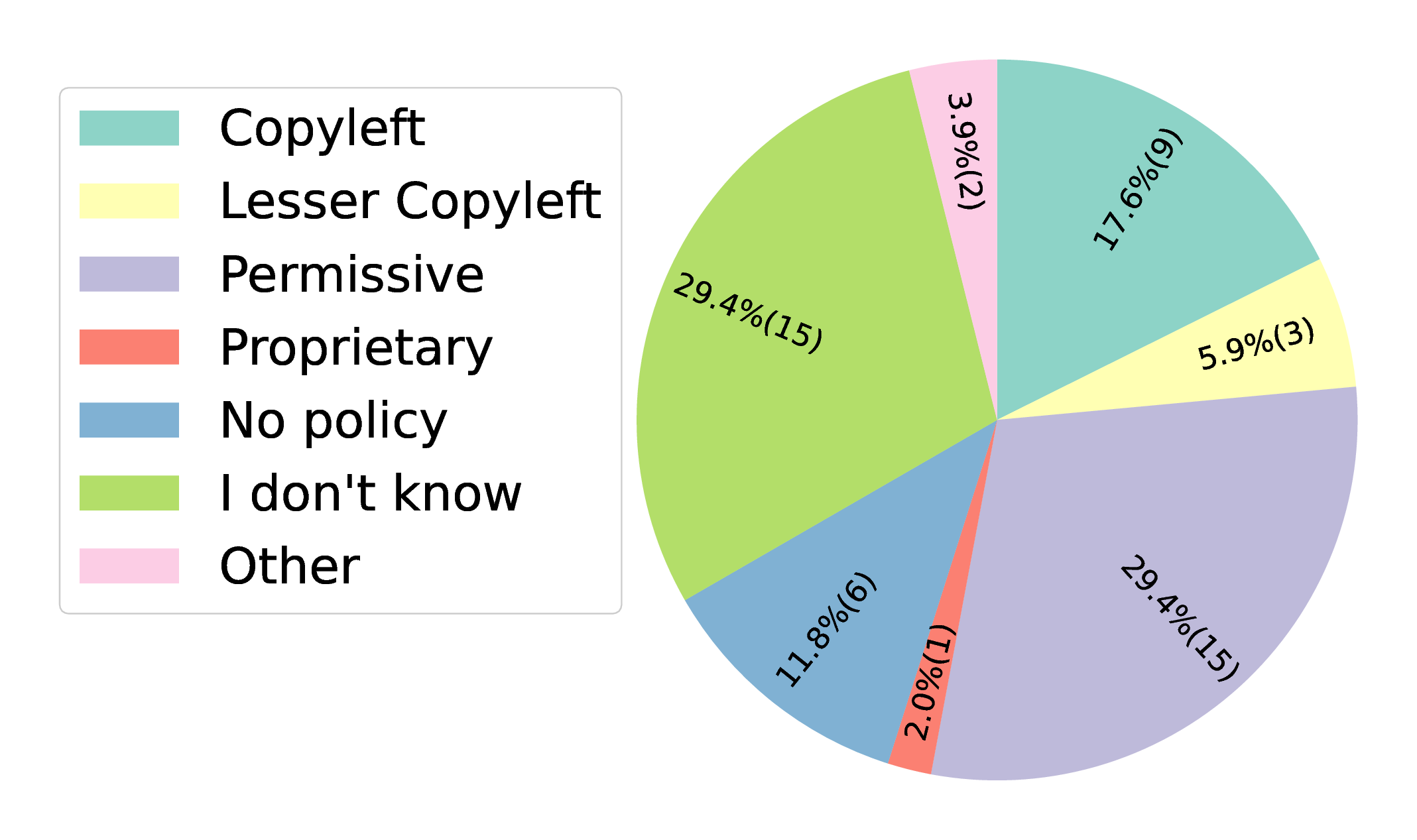}{B10-133_f2}{fig:policy}{Overview of the answers to the question ``What is the default license of your institute/collaboration with respect to software licenses?'' (left) and ``Is the license strongly enforced by the institute or collaboration you are part of?'' (right). Percentages are indicated for each slice, with the corresponding absolute number in parentheses. }

The majority of institutes or collaborations adopted a default license $(54.9\%)$. The majority is distributed over $29.4\%$ (permissive license is the default), $17.6\%$ (copyleft license), $5.9\%$ (lesser copyleft) and $2\%$ (proprietary). Another $29.4\%$ don't know whether a policy is present.

\subsection{Is the license strongly enforced by the institute or collaboration?}
The majority of respondents $(64.7\%)$ say that the license policy is not strongly enforced, while $13.7\%$ tell that the policy is strongly enforced. 

\subsection{What license do you use for your (work related) software?}
\articlefigure[width=8cm]{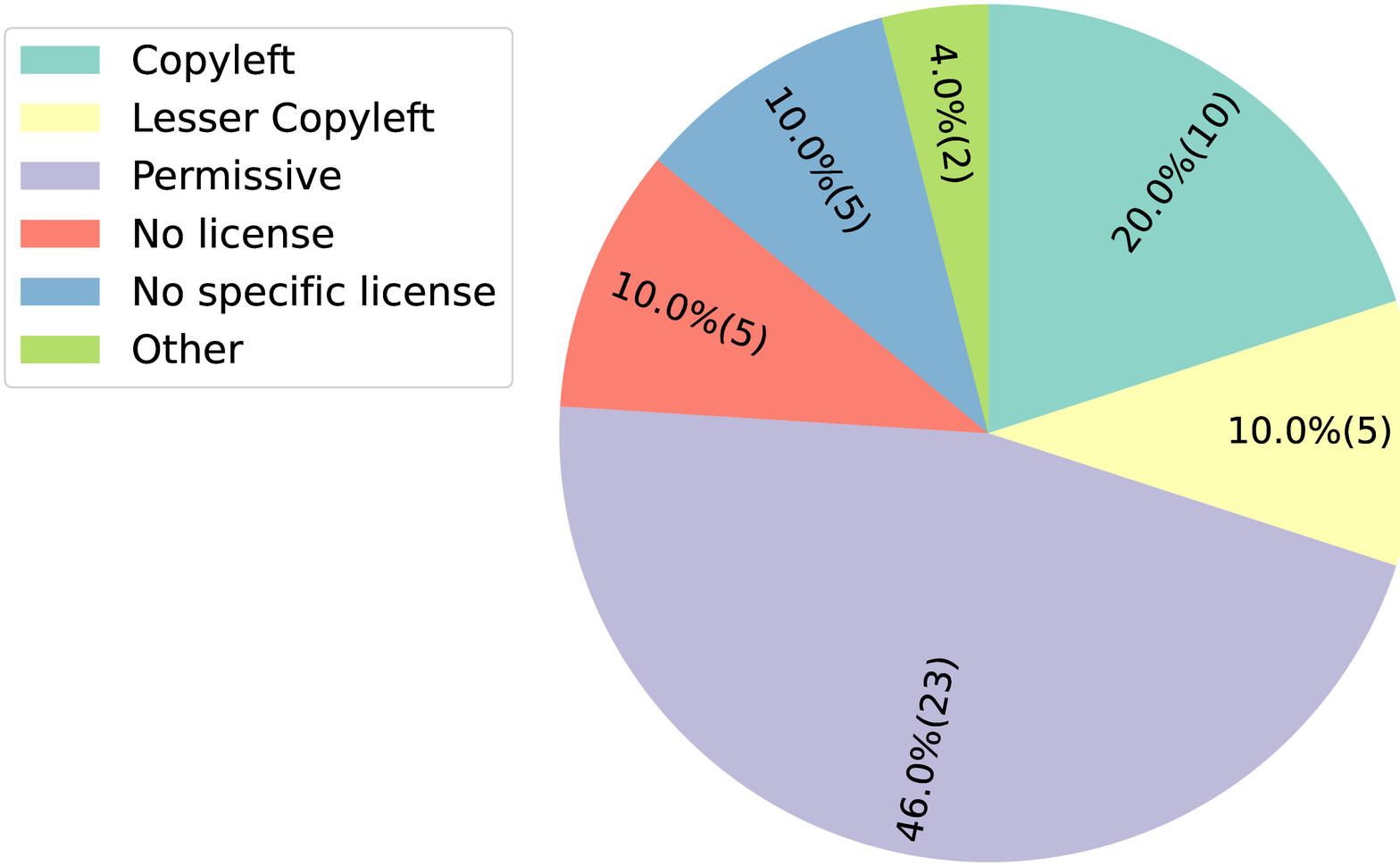}{fig:behaviour}{Overview of the answers to the question ``What license do you use for your (work related) software?''. Percentages are indicated for each slice, with the corresponding absolute number in parentheses.}

\subsection{What license do you use for your (work related) software?}
We added this question as one may know of a policy but still choose not to follow it. The percentages for all types of license are higher than in Fig.~\ref{fig:policy}.  The permissive license stands here out with $46.0\%$. $10\%$ of participants do not use a license at all and another $10\%$ does not use a specific license. These include participants stating they typically only contribute and therefore do not choose a license.

\subsection{Conclusions from the questionnaire}
\subsubsection{Summary}
The results of the questionnaire show that $86\%$ of the participants license their code with open source licenses being the popular ones at about $76\%$. The community prefers permissive licenses at an institutional level ($29.4\%$).  At a personal level ($46.0\%$) are in favour of permissive licenses. (Lesser) copyleft licenses are also popular at $23.5\%$ of institutes and $30.0\%$ of collaborations. A majority of the participants is aware of a policy   $(54.9\%)$, of those $(64.3\%)$ follow it.  A large fraction of participants $(64.7\%)$ say that policies are not strongly enforced.
\subsubsection{Conclusion}
Looking at the numbers where a default license is in place $(54.9\%)$, and people actually applying that license $(35.3\%)$ we conclude that a policy can be effective or at least inspire people to actually apply the advised license.

\section{Main topics of discussion\label{sec:disc}}
The discussion was kicked off with the presentation of two views on the topic of licensing source code:
\begin{itemize}
    \item Australian Astronomical Observatory: Over the years the AAO has moved between licenses during several organisational restructurings.
    \item Cherenkov Telescope Array: Early on in the project the CTA consortium made a very conscious choice for a permissive license.
\end{itemize}

We invited participants to contribute topics for discussion. We asked the attendance to prioritise the five main topics by a vote at the beginning of the session. With 53 votes cast, the most pressing issues were identified (with weight $w$) as
\begin{enumerate}
    \item How to choose a license? $w=0.4$
    \item What to do if licenses in projects/collaborations do not mix? $w=0.3$
    \item What to do if licenses are not respected? $w=0.25$
    \item Good reasons for (open) licenses $w=0.2$
\end{enumerate}

During the discussion the following subjects were identified as being in need of additional information and strategies:
\begin{itemize}
    \item How to work with legacy software without clear licensing?
    \item Who should decide the type of license and how to mention the authors?
    \item How to deal with patents and include companies in software projects?
\end{itemize}

The numbers support that people want to get advice on how to pick the right license for their case, especially in the light that a software publication and citation of it can be helped by a license \citep{B10-129}. It was recognised that the license for a software project should be chosen early because a relicensing of code becomes more time consuming the older a project's source code grows and the more people contribute. This is due to the simple fact that relicensing requires the consent of a source code's copyright holder. In part, this can be avoided by using contributor license agreements to aggregate copyright to a central legal entity.

\section{Prospects for future work\label{sec:forward}}
The main conclusion from the session is that the topic is very relevant to the community and the wish for a common approach exists. As a starting point, we created a knowledge base at \url{https://escape2020.pages.in2p3.fr/wp3/licensing/} which contains the materials presented at the session, slides with additional information and the results of the survey.  It will be expanded in the context of the ESCAPE project.


\bibliography{B10-133}

\end{document}